\documentstyle[11pt,mrs2001,epsfig]{article}
\begin{document}

\title{Clustering of SCUBA galaxies and implications for the Planck mission}  

\author{M.Magliocchetti$^1$, L.Moscardini$^2$, G.de Zotti$^3$,
G.L.Granato$^3$, L.Danese$^1$}
\affil{$^1$SISSA, Via Beirut 4, 34014, Trieste, Italy}
\affil{$^2$Dipartimento di Astronomia, Universit\`a di Padova, Vicolo
dell'Osservatorio 2, I-35122 Padova, Italy}  
\affil{$^3$Osservatorio Astronomico di Padova, Vicolo dell'Osservatorio 5, I-35122
Padova, Italy}

\begin{abstract}
The clustering properties of SCUBA-selected
galaxies are investigated within the framework of a unifying scheme
relating the formation of QSOs and spheroids. The theoretical angular
correlation function is derived for different bias functions,
corresponding to different values of the ratio $M_{\rm halo}/M_{\rm   
sph}$ between the mass of the dark halos hosting such galaxies and the
mass in stars produced at the end of the major star-formation burst.
SCUBA sources are predicted to be strongly clustered, with a
clustering strength increasing with mass. Comparisons with the best available 
measurements show better fits for 
$M_{\rm halo}/M_{\rm sph}\simeq 100$. The model can also
account for the clustering of Lyman-break galaxies, seen as the
optical counterpart of low- to intermediate-mass primeval spheroidal
galaxies. Best agreement is once again obtained for high values of the 
$M_{\rm halo}/M_{\rm sph}$ ratio. We also discuss implications for
small scale fluctuations observed at different wavelengths by
forthcoming experiments such as the Planck mission aimed at mapping the 
Cosmic Microwave Background (CMB).         
\end{abstract}

\section{Clustering of Scuba Sources}

The theoretical expression for the angular two-point correlation function 
$w(\theta)$ can be derived from its
spatial counterpart $\xi$ by projection via the relativistic Limber
equation (Peebles 1980): ~~
\begin{eqnarray}
w(\theta)=2\:\frac{\int_0^{\infty}\int_0^{\infty}N^2(z)\;b_{\rm
eff}^2(M_{\rm min}, z)\;
(dz/dx)\;\xi(r,z)\;dz\;du}{\left[\int_0^{\infty}N(z)\;dz\right]^2},
\label{eq:limber}
\end {eqnarray}
where $x$ is the comoving radial coordinate,
$r=(u^2+x^2\theta^2)^{1/2}$ (for a flat universe and in the small
angle approximation), and $N(z)$ is the number of
objects within the shell ($z,z+dz$).\\
The mass-mass correlation function $\xi(r,z)$ to be inserted in
eq.(\ref{eq:limber}) has been obtained following the work by \cite{12} 
(see also \cite{7} and \cite{mos}), 
which provides an analytical way to derive the trend of $\xi(r,z)$    
both in the linear and non-linear regime. Note that $\xi(r,z)$ {\it only}
depends on the underlying cosmology, which we fix by adopting  
$h_0=0.7$, $\Omega_0=0.3$, $\Lambda=0.7$ and a COBE-normalized value of 
$\sigma_8=1$. The relevant properties of
SCUBA galaxies are included in the redshift distribution of sources $N(z)$, 
and in the bias factor $b_{\rm eff}(M_{\rm min},z)$.
 
The effective bias factor $b_{\rm eff}(M_{\rm min}, z)$ of all the dark matter
haloes with masses greater than some threshold $M_{\rm min}$ is then 
obtained by
integrating the quantity $b(M,z)$ (whose expression has been taken from 
\cite{5}) - representing the bias of
individual haloes of mass $M$ - opportunely weighted by the number
density $n_{SCUBA}(M,z)$ of SCUBA sources:
\begin{eqnarray}                              
b_{\rm eff}(z)=
\frac{\int_{M_{\rm min}}^{\infty} dM\;b(M,z)\;n_{\rm SCUBA}(M,z)}
{\int_{M_{\rm min}}^{\infty} dM\;n_{\rm SCUBA}(M,z)} \ .
\label{eq:bias}
\end{eqnarray}
Note that, as $n_{\rm SCUBA}$ can be thought as the fraction of haloes
hosting a galaxy in the process of forming stars, its expression can
also be written as $n_{SCUBA}(M,z)=n(M,z)\;T_B/t_h$, where $n(M,z)\; dM$ is
the mass spectrum of haloes with masses between $M$ and $M+dM$ (\cite{15}), 
$T_B$ is the duration of
the star-formation burst and $t_h$ is the life-time of the haloes in
which these objects reside (see \cite{9}). 

According to \cite{17}, sources showing up in the SCUBA
counts can be broadly divided into three categories: low-mass (masses
in the range $M_{\rm sph}\simeq 10^9-10^{10}M_\odot$, duration of the
star formation burst $T_B\sim 2$~Gyr, and typical fluxes $S\la
1$~mJy), intermediate-mass ($M_{\rm sph}\simeq
10^{10}$--$10^{11}M_\odot$ and $T_B\sim 1$~Gyr) and high-mass ($M_{\rm
sph}\ga 10^{11}M_\odot$, $T_B\sim 0.5$~Gyr, dominating the counts   
at fluxes $S\ga 5-10$~mJy).  Note that by $M_{\rm sph}$ we denote
the mass in stars at completion of the star formation process.\\ In
order to evaluate the bias factor in eq.(\ref{eq:bias}) we then consider
two extreme cases for the ratio between the mass in stars and the mass
of the host dark halo: $M_{\rm halo}/M_{\rm sph}=100$ and $M_{\rm
halo}/M_{\rm sph}=10$. $M_{\rm halo}/M_{\rm sph}=10$
roughly corresponds to the ratio $\Omega_0/\Omega_{\rm baryon}$
between total and baryon density, where we adopted for the latter
quantity the standard value from primordial nucleosynthesis; this
corresponds to having assumed all the baryons to be locked into stars and, as
a consequence, has to be considered as a conservative lower limit. 
$M_{\rm halo}/M_{\rm sph}=100$ is instead related
to $\Omega_0/\Omega_{\star}$, $\Omega_{\star}$ being the present mass
density in visible stars:
the likely value is expected to be $M_{\rm sph}/M_{\rm halo}=1-3$~\%.
 
Armed with the above results we can then evaluate the two-point
correlation function in eq.~(\ref{eq:limber}) for different $M_{\rm
halo}/M_{\rm sph}$ ratios and different flux cuts. Figure
1 presents our predictions for $w(\theta)$,
respectively for a flux cut of 50 (solid line), 10 (dashed line) and 1
(dotted line) mJy. Higher curves of each kind correspond to the case
$M_{\rm halo}/M_{\rm sph}=100$, while lower curves refer to 
$M_{\rm halo}/M_{\rm sph}=10$.\\
The highest clustering amplitude is found for the brightest sources
($S \ge 50$~mJy). This is because they are associated to the most
massive dark halos and are therefore highly biased tracers of the dark
matter distribution. In addition, according to \cite{17}, they have a rather 
narrow redshift distribution so that the dilution of the clustering signal is
minimum. The very sharp drop of all the curves at $\theta\simeq
1^\circ$ is due to the absence of nearby objects. 
This reflects the notion that the
actively star-forming phase in spheroids is completed at $z> 1$.\\
Note that, since the clustering amplitude strongly depends on the
quantity $M_{\rm
halo}/M_{\rm sph}$, measurements of the angular correlation function
$w(\theta)$ are in principle able to discriminate amongst different
models of SCUBA galaxies and in particular to determine both their
star-formation rate, via the amount of
baryonic mass actively partaking the process of star formation, and the
duration of the star-formation burst.
 
A first attempt to measure the angular correlation function of $S\ge 5$~mJy 
SCUBA sources has been recently presented by \cite{scott}. Although such 
measurements are dominated by noise due to small-number statistics, it is 
nevertheless interesting to note that -- as illustrated by the left-hand panel 
of Figure 2 -- our model (with a preference for the 
$M_{\rm halo}/M_{\rm sph}=100$ case) shows full consistency with the data 
(kindly provided by S. Scott).

\begin{figure}
\vspace{6cm} \includegraphics{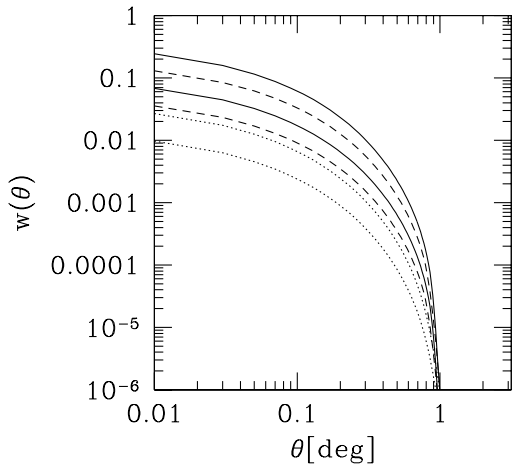}         
\caption{Predictions for the angular correlation function 
$w(\theta)$ of forming spheroids at 850~$\mu$m 
for different halo-to-bulge mass ratios and different flux cuts.
Solid lines are for sources brighter than 50~mJy, dashed lines for
sources brighter than 10~mJy, while dotted lines correspond to $S\ge
1$~mJy. Higher curves of each type correspond to $M_{\rm halo}/M_{\rm
sph}=100$, lower ones to $M_{\rm halo}/M_{\rm sph}=10$.} 
\end{figure}

\begin{figure}
\vspace{6cm} 
\includegraphics{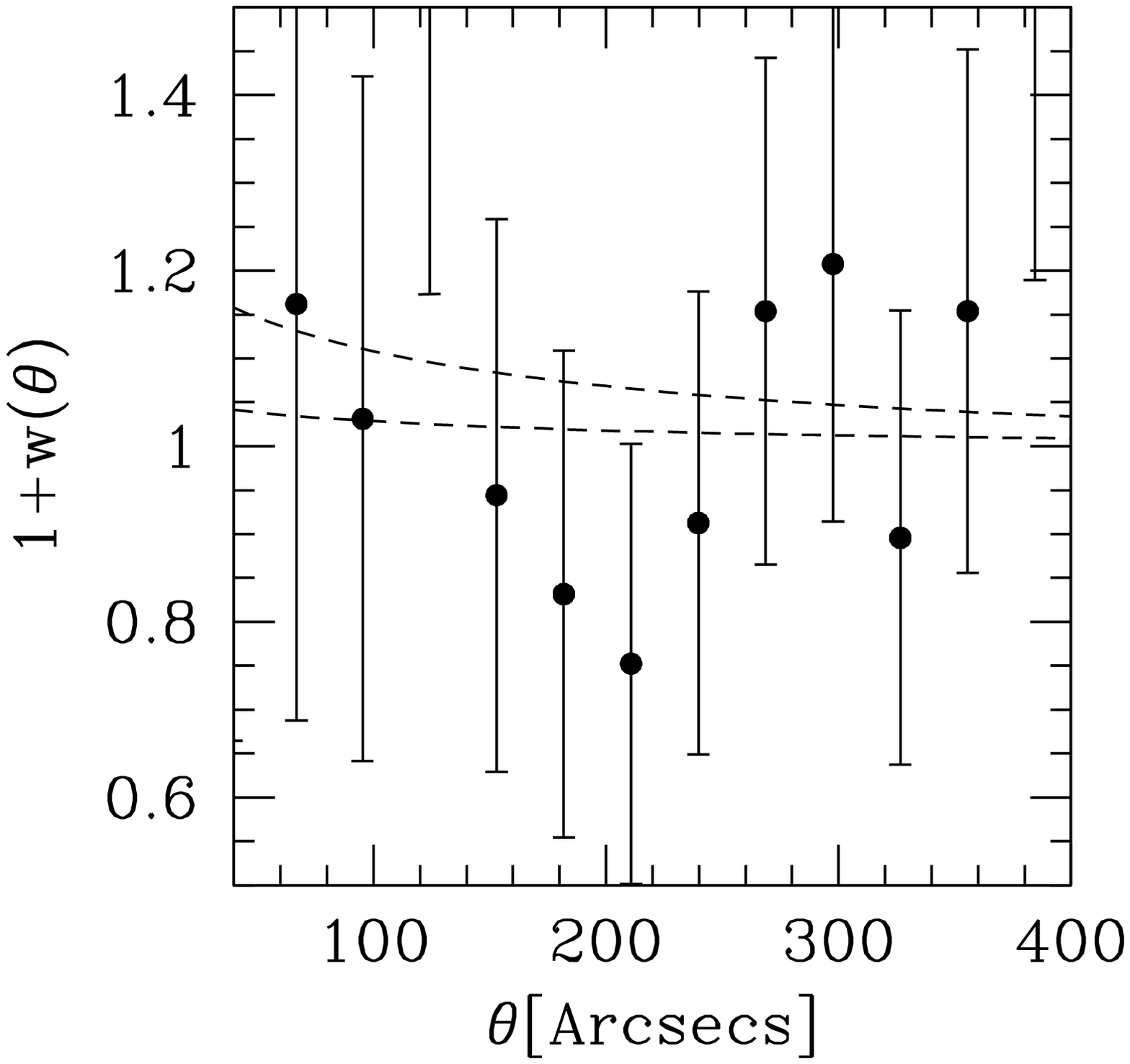}
\includegraphics{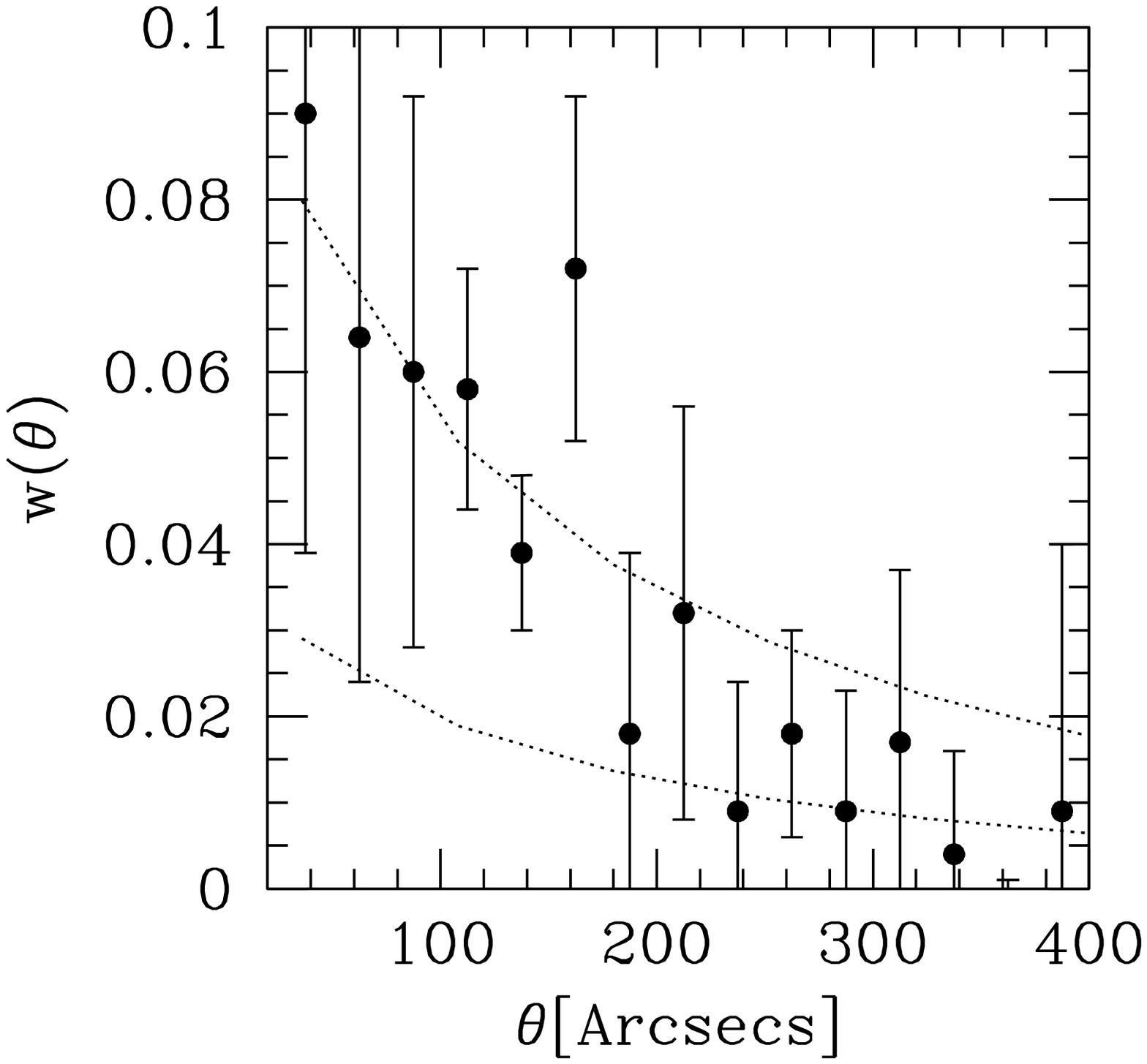} 
\caption{Left-hand panel: angular correlation function of $S\ge 5$~mJy 
SCUBA sources. Model predictions are
illustrated by the dashed lines (where the higher one is for $M_{\rm halo}
/M_{\rm sph}=100$ and the lower one for $M_{\rm halo}/M_{\rm sph}=10$), 
while data points represent the \cite{scott} measurements. 
Right-hand panel: angular correlation function of LBGs. Model 
predictions are
shown by the dotted lines (the higher one for 
$M_{\rm halo}/M_{\rm sph}=100$, the lower one for 
$M_{\rm halo}/M_{\rm sph}=10$), while data
points represent the \cite{3} measurements.}
\end{figure}

Another possible way to obtain some information on the nature of SCUBA sources 
via their clustering properties is provided by the predictions of \cite{17} 
for Lyman-break galaxies to be the low- to intermediate-mass tail of primeval 
spheroidal galaxies, with
$T_B \sim 1-2$~Gyr and a star-formation rate ranging from a few to a
hundred $M_\odot\; \hbox{yr}^{-1}$.  In
Figure 2 (right-hand panel) we then plotted the predicted $w(\theta)$ for 
those sources with $S\ge 1$~mJy (corresponding to $M_{\rm sph} \ga 10^{10}
M_\odot$), expected to be found within the redshift range $2.5 \le z \le 3.5$, 
covered by the original (Steidel et al., 1996) sample. As in the former case, 
the higher curve is for $M_{\rm halo}/M_{\rm sph}=100$, the lower one for
$M_{\rm halo}/M_{\rm sph}=10$. The data points show the \cite{3} 
measurements. Even though large errors once again affect the observational 
findings, it is nevertheless clear that
the predicted trend for $w(\theta)$ can correctly reproduce the data for 
high ($80-100$) values of the $M_{\rm halo}/M_{\rm sph}$ ratio. This result
is consistent with the predictions by \cite{17} and implies a well 
defined relationship between SCUBA galaxies and LBGs. Furthermore, it 
also confirms the expectations for a
small fraction (on the order of a few percent) of the total mass to be
confined into stars.

Finally, it is also worth noticing that our predictions are in agreement with 
the strong clustering of EROs recently detected by \cite{1}, since 
we expect (\cite{17}) these objects to be the direct descendants of 
SCUBA galaxies, and therefore to exhibit the same 
clustering properties.  
                                                                
\section{Power Spectrum of Temperature Fluctuations}

An issue intimately connected with the analysis of galaxy clustering
is the study of the contribution of
unresolved sources (i.e. sources with fluxes fainter than some
detection limit $S_d$) to the background intensity. Its general
expression is given by:
\begin{eqnarray}
I=\int_0^{S_d}
\frac{dN}{dS}\;S\;dS=\frac{1}{4\pi}\int^{L_{\rm max}}_{L_{\rm min}}
d{\rm log}L\;L\;\int_{z(S_d,L)}^{z_{\rm max}}\!\!\!\!\!
dz\;\Phi(L,z)\;\frac{K(L,z)}{(1+z)^2}\;\frac{dx}{dz}
\label{eq:I}
\end{eqnarray}
(see e.g. \cite{dezo}), where $dN/dS$ denotes the
differential number counts, $L_{\rm max}$ and $L_{\rm min}$ are
respectively the maximum and minimum local luminosity of the sources,
$K(L,z)$ is the K-correction, $z_{\rm max}$ is
the redshift when the sources begin to shine, $z(S_d,L)$ is the
redshift at which a source of luminosity $L$ is seen with a flux equal
to the detection limit $S_d$, $\Phi(L,z)$ is the luminosity function
(i.e. the comoving number density of sources per unit $d{\rm log}L$),
and $x$ is the comoving radial coordinate.

The intensity fluctuation $\delta I$ due to inhomogeneities in the
space distribution of unresolved sources is then given by
eq. (\ref{eq:I}), with the quantity $\Phi(L,z)$ replaced by $\delta
\Phi(L,z)$. It is easily shown that the angular correlation of such
intensity fluctuations
$C(\theta)=\langle \delta I(\theta^\prime, \phi^\prime)\; \delta
I(\theta'', \phi'')\rangle$,
where $(\theta^\prime, \phi^\prime)$ and $(\theta'', \phi'')$ define
two positions on the sky separated by an angle $\theta$, can be
expressed as the sum of two terms $C_P$ and $C_C$, the first one due
to Poisson noise (i.e. fluctuations given by randomly distributed
objects), and the second one owing to source clustering. It is possible to 
show (\cite{8}) that, in the case of highly clustered sources, the Poissonian 
term $C_P$ is negligible with respect to the one due to clustering. In the 
following we therefore only
concentrate on temperature fluctuations caused by the $C_C$ term
(hereafter simply called $C$). 

By making use of the quantities previously defined and of
eq.~(\ref{eq:I}), the clustering term $C$ takes the form:
\begin{eqnarray}
C(\theta)=\left({1\over 4\pi}\right)^2  \int_{z_{(L_{\rm
min},S_d})}^{z_{\rm max}}dz\;b_{\rm eff}^2(M_{\rm min},z)\; \frac{j^2_{\rm eff}(z)}
{(1+z)^4}\left(\frac{dx}{dz}\right)^2\int_0^\infty du\;\xi(r,z),
\label{eq:cth}
\end{eqnarray}
where the effective volume emissivity $j_{\rm eff}$ is expressed as:
\begin{eqnarray}
j_{\rm eff}=\int_{L_{\rm min}}^{{\rm min}[L_{{\rm
max},L(S_d,z)}]}\!\!\!\!\!  \Phi(L,z)\; K(L,z)\; L\;d{\rm log}L
\label{eq:jeff}
\end{eqnarray}.

$C(\theta)$ in eq.(\ref{eq:cth}) has been evaluated 
separately for the three cases of low-, intermediate- and high-mass objects, 
by plugging in
eq.~(\ref{eq:jeff}) the appropriate expressions for the luminosity
function. The total contribution of clustering to
intensity fluctuations has then been derived by adding up all the values of 
$C(\theta)$ obtained for the different mass intervals and
by also taking into account the cross-correlation terms between objects
of different masses, according to the expression
\begin{eqnarray}
C^{TOT}(\theta)=\sum_{i,j=1}^3 {\sqrt {C_i(\theta)C_j(\theta)}},
\label{eq:ctot}
\end{eqnarray}
where the indexes $i$,$j$ stand for high, intermediate and low masses.\\
Note that, the quantity $b_{\rm eff}^2(M_{\rm min},z)$ in eq.(\ref{eq:cth})
should indeed be read as $b_{\rm eff}^2(M_{\rm min},M_{\rm max},z)$, where
$M_{\rm max}$ is the maximum halo mass corresponding to the maximum
visible bulge mass (i.e. upper limit for the mass locked into stars), which
 corresponds in eq.(\ref{eq:bias}) to a replacement in the upper limit 
of the integrals of $\infty$ with $M_{\rm max}$.

\begin{figure}
\vspace{6cm} \includegraphics{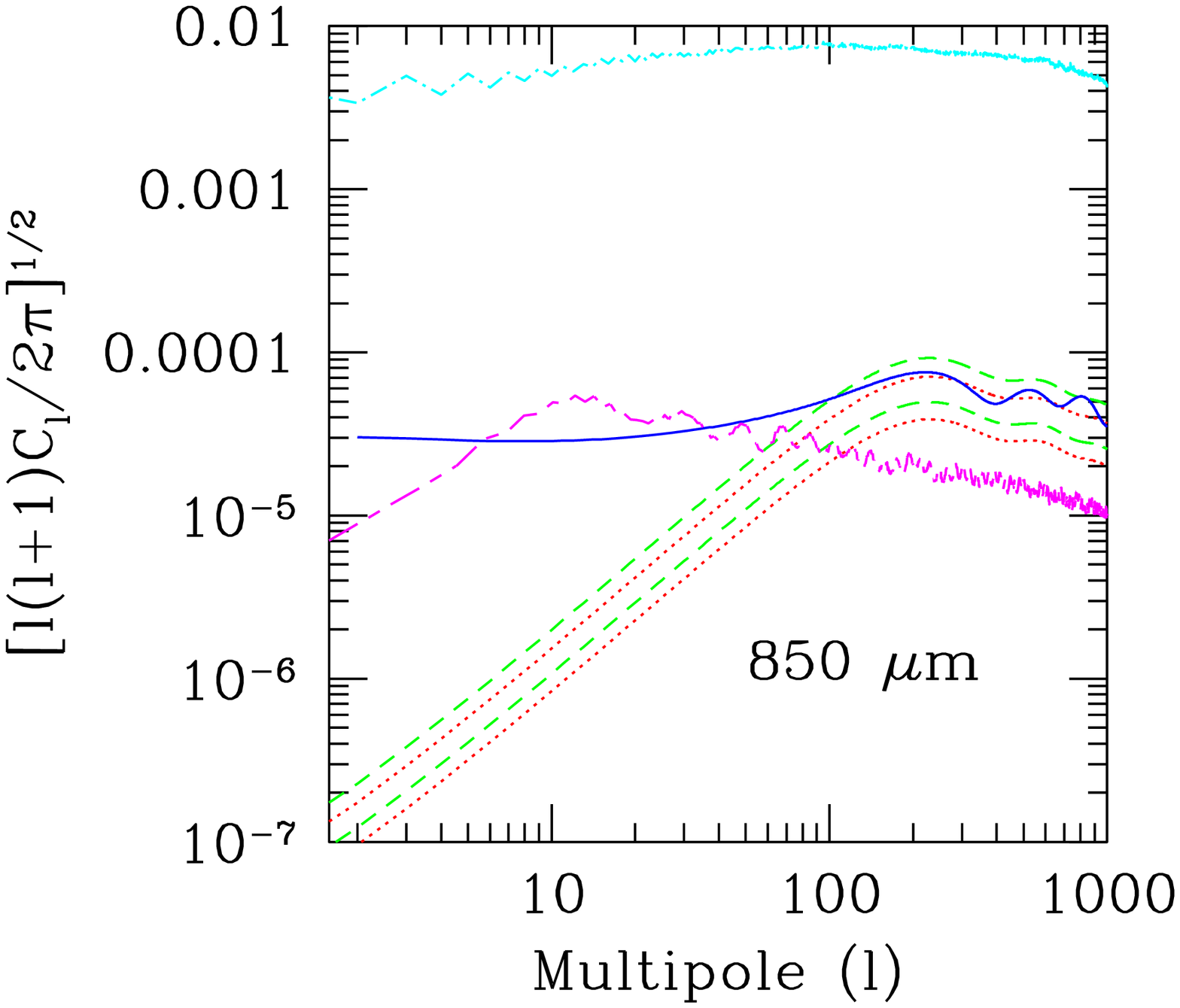}         
\includegraphics{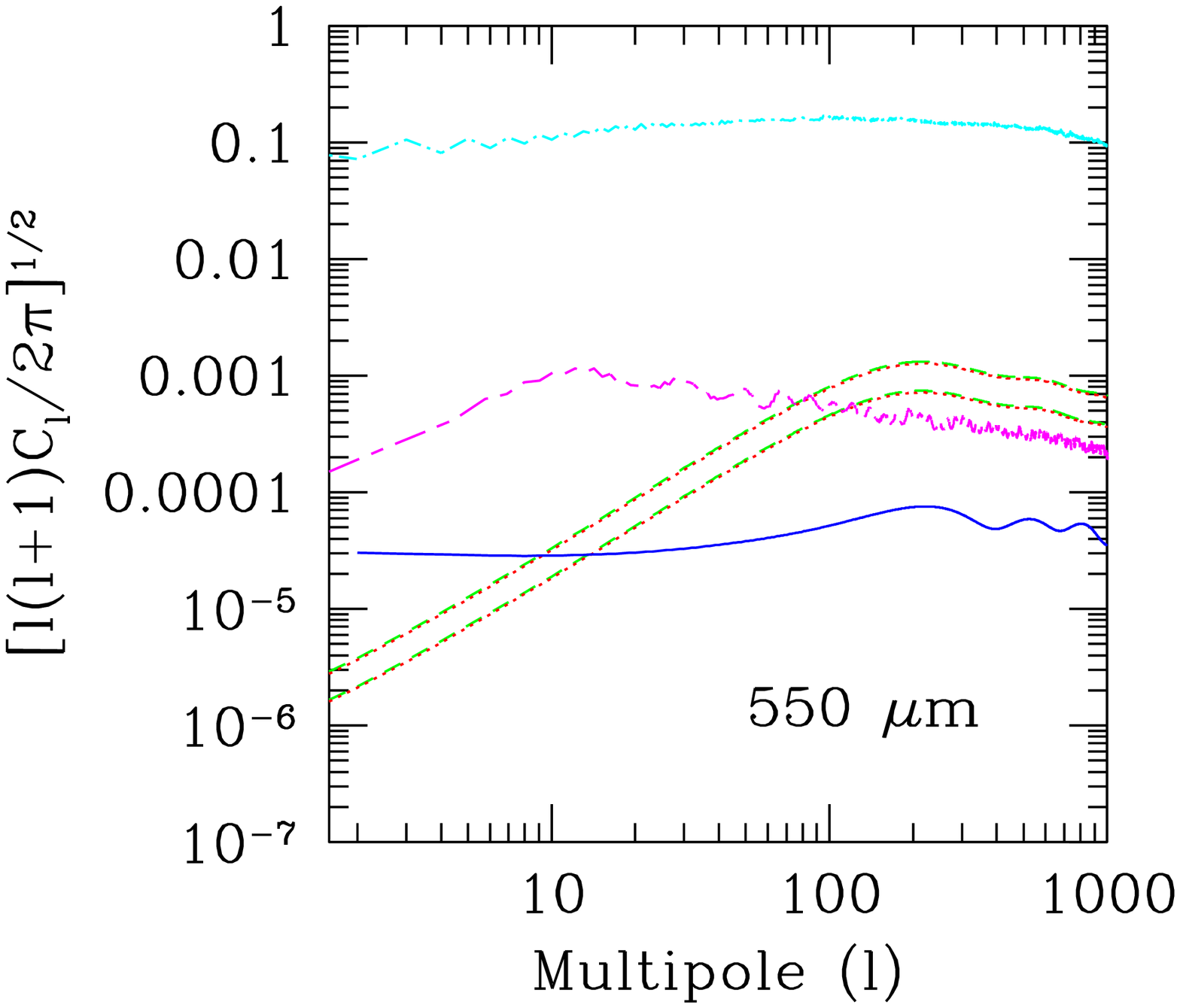}
\caption{Predicted power spectrum of temperature fluctuations $\delta
T_l=\sqrt{l(l+1)C_l/2\pi}$ (in units of K) as a function of the
multipole $l$ at 353 GHz (left-hand panel) and 545 GHz (right-hand panel), 
the central frequencies of two Planck/HFI
channels.  Left-hand panel: dashed lines are for a detection limit 
$S_d=100$~mJy,
dotted ones for $S_d=10$~mJy. In both cases higher curves are
obtained for $M_{\rm halo}/M_{\rm sph}$=100, lower ones for
$M_{\rm halo}/M_{\rm sph}$=10. The solid line represents the power
spectrum of primary CMB anisotropies as predicted by a standard Cold
Dark Matter model for a $\Lambda$CDM cosmology ($\Lambda=0.7$,
$\Omega_0=0.3$, $h_0=0.7$). Right-hand panel: as in the former case but with 
flux detection limits of 450 (dashed lines) and 45 (dotted lines) mJy. 
The overlap observed here is due to paucity of bright ($S\ga 50$~mJy) 
sources. In both 
panels the upper dashed-dotted curve represents the contribution from galactic 
dust emission averaged all over the sky, while the lower one illustrates the 
case obtained by restricting the analysis to high galactic latitudes 
($80\deg\le b\le 90\deg$). 
\label{fig:dT}}
\end{figure}

The angular power spectrum of the intensity fluctuations
can then be obtained via (see \cite{dezo}):
\begin{eqnarray}
\delta T(\theta)=\langle\left(\Delta T\right)^2
\rangle^{1/2} =\frac{\lambda^2\!
\sqrt{C^{TOT}(\theta)}}{2\;k_b}\left[{\rm exp}\left (\frac{h\nu}{k_b
T}\right)-1 \right]^2{\rm exp}\left(- \frac{h\nu}{k_b
T}\right)/\left(\frac{h\nu}{k_b T} \right)^2,
\end{eqnarray}
which relates intensity and temperature fluctuations.
Figure \ref{fig:dT} shows the predicted values for the quantity
$\delta T_l=\sqrt{l(l+1)C_l/2\pi}$ (in units of K) at respectively 353 GHz
($850\,\mu$m -- left-hand panel) and 545 GHz ($550\,\mu$m -- right-hand 
panel) -- the central frequencies of two of the channels of the High Frequency 
Instrument (HFI) 
of the ESA's Planck mission -- as a function of the multipole $l$ up to 
$l=1000$. Results are plotted for
two different values of the source detection limit ($S_d=100$ and 10~mJy 
for the 353 GHz case and $S_d=450$ and 45~mJy for the 545 GHz case) and the 
usual two values of $M_{\rm halo}/M_{\rm sph}$.  
Also shown, for comparison, is the power spectrum of primary
(CMB) anisotropies (solid line) predicted for the cosmology specified in the 
caption, computed with the CMBFAST code developed by \cite{sel}.

At $850\,\mu$m, our model predicts fluctuations of amplitude due to
clustering comparable to (and possibly even larger than) those
obtained for primary CMB anisotropies at $l\ga 50$. This is because 
most of the clustering signal comes from
massive galaxies with fluxes $S\ga 10$~mJy, which lie at substantial
redshifts and are therefore highly biased
tracers of the underlying mass distribution.  Also the strongly
negative K-correction increases their contribution to the effective
volume emissivity [eq.~(\ref{eq:jeff})] and therefore to the
fluctuations. The $550\,\mu$m case is even more striking, since the 
contribution from clustering is expected to be more than an order of 
magnitude greater than the one originating from primordial fluctuations, 
regardless of the flux detection limit.\\
This implies that important information on the clustering properties of faint
sub-mm galaxies (and hence on physical properties such as their
mass and/or the amount of baryons involved in the star-formation
process) will reside in the Planck maps at frequencies greater than 353 GHz 
where, however, the dominant signal is expected to come from interstellar dust
emission. In order to show this effect, in Figure 3 we have also plotted the 
expected contribution from galactic dust emission averaged all over the sky 
(upper dashed-dotted curves). This was derived from the IRAS maps at 
$60\,\mu$m, rescaled at the frequencies under exam by assuming a 
grey-body 
spectrum with $T_{DUST}=18$~K. As already anticipated, this signal appears to 
be the dominant one in both the 353 and 545 GHz Planck channels. 
Nevertheless, it is
still possible to extract information on the nature of sub-mm galaxies if one  
restricts
the analysis to high galactic latitude regions (i.e. $80\deg\le b\le 90\deg$, 
lower dashed-dotted curves in Figure 3), which are the least affected by 
galactic dust emission. In fact, as it can be seen from Figure 3, the dust 
contribution in this region becomes less important than the one due to the  
clustering of unresolved sources for $l\ga 80$. 
     
\vfill

\begin{thebibliography}{}{
\bibitem{1} Daddi, E., Cimatti, A., Pozzetti, L., Hoekstra, H., R\"ottgering,
H.J.A., Renzini,
A., Zamorani, G., Mannucci, F. 2000, A\&A 361, 535
\bibitem{dezo}
De Zotti, G., Franceschini, A., Toffolatti, L., Mazzei, P., Danese, L., 1996,
Ap. Lett.Comm. 35, 289    
\bibitem{3} Giavalisco,
M., Steidel, C.C., Adelberger, K.L.,  Dickinson, M.E., Pettini, M.,
Kellogg, M. 1998, ApJ 503, 543
\bibitem{17} Granato, G.L., Silva, L., Monaco, P., Panuzzo, P., Salucci, P.,
De Zotti, G., Danese, L. 2001, MNRAS 324,757
\bibitem{5} Jing, Y.P. 1998, ApJ 503, L9
\bibitem{7} Magliocchetti, M., Bagla, J., Maddox, S.J., Lahav, O. 2000,
MNRAS 314, 546                       
\bibitem{8} Magliocchetti, M., Moscardini, L.,
        Panuzzo, P., Granato, G.L., De Zotti, G., Danese, L. 2001, MNRAS,
accepted, astro-ph/0102464
\bibitem{9} Martini, P. \& Weinberg, D.H., 2001, ApJ 547, 12
\bibitem{mos}
Moscardini L., Coles P., Lucchin F., Matarrese S., 1998, MNRAS, 299, 95
\bibitem{12} Peacock, J.A. \& Dodds S.J., 1996, MNRAS 267, 1020
\bibitem{13} Peebles, P.J.E. 1980, The Large-Scale Structure of the Universe,
Princeton University Press
\bibitem{sel} Seljak, U., Zaldarriaga, M. 1996, ApJ  469, 437.
\bibitem{scott} Scott, S.E., et al., astro-ph/0107446
\bibitem {15}Sheth, R.K. \& Tormen, G. 1999, MNRAS 308, 119
\bibitem {16}Smail, I., Ivison, R.J., Blain, A.W. 1997, ApJ 490, L5
}
\end{thebibliography}
\end{document}